\documentclass[12pt]{article}
\usepackage{graphicx}
\usepackage{epstopdf}

\def\beq{\begin{equation}}
\def\eeq{\end{equation}}
\def\to{\rightarrow}

\def\bsg{\ifmmode B\to X_s\gamma\else $B\to X_s\gamma$\fi}
\def\bsll{\ifmmode B\to X_s\ell^+\ell^-\else $B\to X_s\ell^+\ell^-$\fi}
\def\bstt{\ifmmode B\to X_s\tau^+\tau^-\else $B\to X_s\tau^+\tau^-$\fi}
\def\shat{\ifmmode \hat{s}\else $\hat{s}$\fi}

\newcommand{\newc}{\newcommand}

\newc{\lcal}{\int {\cal L}dt}

\newc{\LSP}{{\chi^0_1}}
\newc{\stauR}{{\tilde \tau_R}}
\newc{\stau}{{\tilde \tau_1}}
\newc{\mstop}{m_{\tilde{t}}}
\newc{\mHpm}{m_{H^\pm}}
\newc{\gsim}{\lower.7ex\hbox{$\;\stackrel{\textstyle>}{\sim}\;$}}
\newc{\lsim}{\lower.7ex\hbox{$\;\stackrel{\textstyle<}{\sim}\;$}}
\newc{\ie}{{\it i.e.}}
\newc{\etal}{{\it et al.}}
\newc{\eg}{{\it e.g.}}
\newc{\kev}{\hbox{\rm\,keV}}
\newc{\mev}{\hbox{\rm\,MeV}}
\newc{\gev}{\hbox{\rm\,GeV}}
\newc{\tev}{\hbox{\rm\,TeV}}
\newc{\xpb}{\hbox{\rm\, pb}}
\newc{\xfb}{\hbox{\rm\, fb}}

\newc{\mtop}{m_t}
\newc{\mbot}{m_b}
\newc{\mz}{M_Z}
\newc{\mw}{M_W}
\newc{\alphasmz}{\alpha_s(M_Z)}
\newc{\swsq}{\sin^2\theta_W}
\newc{\cwsq}{\cos^2\theta_W}
\newc{\tw}{\tan\theta_W}
\newc{\cw}{\cos\theta_W}
\newc{\sw}{\sin\theta_W}
\newc{\BR}{\hbox{\rm BR}}
\newc{\zbb}{Z\to b\bar}
\newc{\Gb}{\Gamma (Z\to b\bar b)}
\newc{\Gh}{\Gamma (Z\to \hbox{\rm hadrons})}
\newc{\rbsm}{R_b^\hbox{\rm sm}}
\newc{\rbsusy}{R_b^\hbox{\rm susy}}
\newc{\drb}{\delta R_b}

\newc{\sgn}{\mbox{sgn}}

\def\eq#1{eq.~(\ref{#1})}

\def\beqa{\begin{eqnarray}}
\def\eeqa{\end{eqnarray}}

\def\mtil{\tilde{m}}
\def\bino{\tilde{B}}
\def\wino{\tilde{W}}
\def\hino{\tilde{H}}

\def\beq{\begin{equation}}
\def\eeq{\end{equation}}
\def\bea{\begin{eqnarray}}
\def\eea{\end{eqnarray}}
\def\slashchar#1{\setbox0=\hbox{$#1$}           
   \dimen0=\wd0                                 
   \setbox1=\hbox{/} \dimen1=\wd1               
   \ifdim\dimen0>\dimen1                        
      \rlap{\hbox to \dimen0{\hfil/\hfil}}      
      #1                                        
   \else                                        
      \rlap{\hbox to \dimen1{\hfil$#1$\hfil}}   
      /                                         
   \fi}                                         %
\catcode`@=11
\long\def\@caption#1[#2]#3{\par\addcontentsline{\csname
  ext@#1\endcsname}{#1}{\protect\numberline{\csname
  the#1\endcsname}{\ignorespaces #2}}\begingroup
    \small
    \@parboxrestore
    \@makecaption{\csname fnum@#1\endcsname}{\ignorespaces #3}\par
  \endgroup}
\catcode`@=12

\begin{document}

\baselineskip=18pt

\setcounter{footnote}{0}
\setcounter{figure}{0}
\setcounter{table}{0}

\begin{titlepage}
January 2006 \hspace*{\fill}
CERN-PH-TH/2005-260

\begin{center}
\vspace{1cm}

{\Large \bf The Well-Tempered Neutralino}

\vspace{0.8cm}

{\bf N. Arkani-Hamed$^1$, A. Delgado$^2$, G.F. Giudice$^2$}

\vspace{.5cm}

$^1${\it Jefferson Laboratory of Physics, Harvard University,\\
         Cambridge, Massachusetts 02138, USA}

$^2${\it CERN, Theory Division, CH-1211 Geneva 23, Switzerland}

\end{center}
\vspace{1cm}

\begin{abstract}
\medskip
The dark-matter prediction is usually considered as one of the
successes of low-energy supersymmetry. We argue that, after LEP
constraints are taken into account, the correct prediction for the
dark-matter density, at a quantitative level, is no longer a natural
consequence of supersymmetry, but it requires special relations among
parameters, highly sensitive to small variations. This is analogous to
the problem of electroweak-symmetry breaking, where the correct value
of the $Z$ mass is obtained only with a certain degree of fine
tuning. In the general parameter space of low-energy supersymmetry,
one of the most plausible solution to reproduce the correct value of
the dark-matter density is the well-tempered neutralino, which
corresponds to the boundary between a pure Bino and a pure Higgsino or
Wino. We study the properties of well-tempered neutralinos and we
propose a simple limit of split supersymmetry that realizes this situation.

\end{abstract}

\bigskip
\bigskip

\end{titlepage}


\section{The Supersymmetric Dark Matter Impasse}
\label{sec1}

The natural prediction of thermal-relic dark matter is usually considered as
one of the most attractive features of models with low-energy supersymmetry.
Indeed a stable, neutral, colourless, weakly-interacting particle with
Fermi-scale mass leads to a present dark-matter density in rough agreement
with observations~\cite{wmap,pdg}
\beq
\Omega_{\rm DM}h^2=0.113\pm 0.009,
\label{dmobs}
\eeq
quite independently of any detail of the cosmological
evolution at temperatures higher than the freeze-out temperature
$T_f\sim 1$--$100\gev$. Low-energy supersymmetry with conserved R-parity
gives a satisfactory theoretical framework for the existence of such a
particle~\cite{gold}.

Here we want to argue that, although dark matter was certainly a natural
prediction of supersymmetry in the pre-LEP epoch, it is generically
no longer true, at a quantitative level, after LEP data are taken into
account. To illustrate the problem, let us consider the supersymmetric
extension of the Standard Model with minimal field content, and
with general soft terms. An acceptable thermal dark-matter
candidate is obtained
in the case in which a neutralino is the lightest supersymmetric
particle (LSP). As LEP data are forcing the soft terms to be typically
larger
than $M_Z$, a description of the neutralino in terms of current
eigenstates is becoming increasingly appropriate. We will then start
our discussion
with the case in which the neutralino is pure Bino, Wino or Higgsino
and later generalize to mixed states.

\subsection{Bino}

The Bino is a gauge singlet whose
annihilation in the early
universe occurs through squark and slepton exchange.
Since sleptons are usually lighter than squarks and right-handed sleptons
have the largest hypercharge, the Bino annihilation
cross section and its contribution to the present
$\Omega$ are well approximated by
\beq
\langle \sigma_{\bino} v\rangle
= \frac{3g^4\tan^4\theta_Wr(1+r^2)}{2\pi m_{{\tilde e}_R}^2x(1+r)^4}
, ~~~x\equiv \frac{M_1}{T}
,~~~r\equiv \frac{M_1^2}{m_{{\tilde e}_R}^2},
\eeq
\beq
\Omega_{\bino}h^2=1.3\times 10^{-2}\left( \frac{m_{{\tilde e}_R}}{100\gev}
\right)^2 \frac{(1+r)^4}{r(1+r^2)}\left( 1+0.07\log\frac{\sqrt{r}100\gev}
{m_{{\tilde e}_R}} \right) .
\label{ombi}
\eeq
Here $M_1$ is the Bino mass and $m_{{\tilde e}_R}$ is the mass of any of the
three degenerate right-handed sleptons.
The value of $\Omega_{\bino}h^2$ as a function of $M_1$ is shown in
fig.~\ref{fig1} for $M_1/m_{{\tilde e}_R}$ varying between $0.9$ and $0.3$.
For $r$ very close to 1, co-annihilation
becomes important, and \eq{ombi} is no longer valid. We will discuss
co-annihilation later, and for the moment we assume $M_1/m_{{\tilde e}_R}<0.9$.
Then, requiring
that dark matter is constituted by relic Binos, we find that \eq{dmobs}
implies $m_{{\tilde e}_R}<111\gev$ at 95\% CL.
Present LEP limits~\cite{lep} require
$m_{{\tilde e}_R}>100\gev$,
$m_{{\tilde \mu}_R}>97\gev$,
$m_{{\tilde \tau}_R}>93\gev$ at 95\% CL (for neutralino mass of 40~GeV).
Therefore, while before LEP a pure Bino appeared as
an adequate dark-matter candidate in broad
range of parameters, after LEP the typical prediction is that the Bino
relic density is too large.

As it is well known, agreement with dark-matter
observation can still be achieved
under special conditions on the supersymmetry soft terms. One well-studied
possibility~\cite{stau} is that the Bino and the
lightest slepton (usually the stau)
are nearly degenerate in mass,
so that co-annihilation~\cite{coann} reduces the relic abundance. This requires
a rather precise correlation among parameters,
$(m_{\tilde \tau}-M_1)/M_1\lsim T_f/M_1\sim 5\%$. Another possibility
to reduce the relic abundance is to exploit an almost-resonant $s$-channel
annihilation~\cite{reson}.
This requires $|m_H -2 M_1|/m_H\ll 1$, where $m_H$ is the
mass of the CP-odd or the heavy CP-even Higgs boson. The occurrance of
resonant light-Higgs exchange is almost ruled out by data.

There are also alternative ways to reduce the Bino relic density, which
however depend somehow on the
cosmological history. If the reheat temperature
after a period of entropy
production $T_{RH}$ is lower than the freeze-out
temperature $T_f$, than the actual dark-matter
relic density can be much smaller than what is estimated by the usual
thermal calculation~\cite{lowr}.
In this case, the final abundance has a steep dependence
on $T_{RH}$.

A different possibility is to assume that the Bino is only the
next-to-lightest supersymmetric particle (NLSP), allowing for its decay. If
the true LSP is sufficiently weakly coupled to have a negligible thermal relic
abundance, but strongly-enough coupled for the Bino decay to be
cosmologically harmless, then $\Omega_{\rm LSP} =(m_{\rm LSP}/M_1)
\Omega_{\bino}$ can be reconciled with data, for an appropriate
value of $m_{\rm LSP}$. The gravitino is the
most motivated candidate for this scenario~\cite{grav,lesz},
but the more exotic cases of
axino or modulino have also been considered~\cite{axino}.
A difficulty often encountered
in this approach
is that the late Bino decay can upset the usual nucleosynthesis predictions.
Take the case of a gravitino with mass $m_{\tilde G}$.
If we require that the Bino
decay occurs before the onset of nucleosynthesis, then $m_{\tilde G}
<(M_1/100\gev )^{5/2}~10\mev$. To obtain a sufficient $\Omega_{\tilde G}$
from Bino decay with such a low value of $m_{\tilde G}$, it is necessary to
have $m_{{\tilde e}_R}>2\tev (M_1/100\gev )^{1/8}$.
Otherwise the impact of the decay on nucleosynthesis has to be carefully
studied, and indeed the case of a NLSP
neutralino is excluded~\cite{grav}. The situation is better for
a $\tilde \tau$ NLSP~\cite{grav}, but a recent analysis~\cite{lesz}
claims that, in the context of the minimal supergravity model,
a dark-matter gravitino
coming from $\tilde \tau$ decay is also excluded.
A way out is to rely on a
thermal population of gravitinos,
but then the relic-abundance prediction is sensitive on initial cosmological
conditions, through $T_{RH}$.

\subsection{Higgsino}

The Higgsino is traditionally considered a less  favourable dark-matter
candidate,
because it is not the LSP in constrained supersymmetric models  and because its
annihilation cross section is very efficient. The first argument is very
much tied to the assumption of universality where a large Higgsino mass
$\mu$ is necessary
to cancel the positive contribution to $M_Z^2$ from all other soft terms, in
the radiative electroweak mechanism. Smaller values of $\mu$ occur in the
so-called hyperbolic branch~\cite{hyp} and focus point~\cite{focus}.
However, as soon as we
depart from universality, the request of large $\mu$ is no longer justified.
Consider for instance the case in which the soft mass terms
for the stops and for
the Higgses at the cut-off scale are different. Since, for sufficiently
high cut-off, the stops give a positive
contribution to $M_Z^2$ and the Higgses give a negative contribution, an
appropriate cancellation can be obtained for any value of $\mu$.

\begin{figure}
\centering
\includegraphics[width=0.8\linewidth]{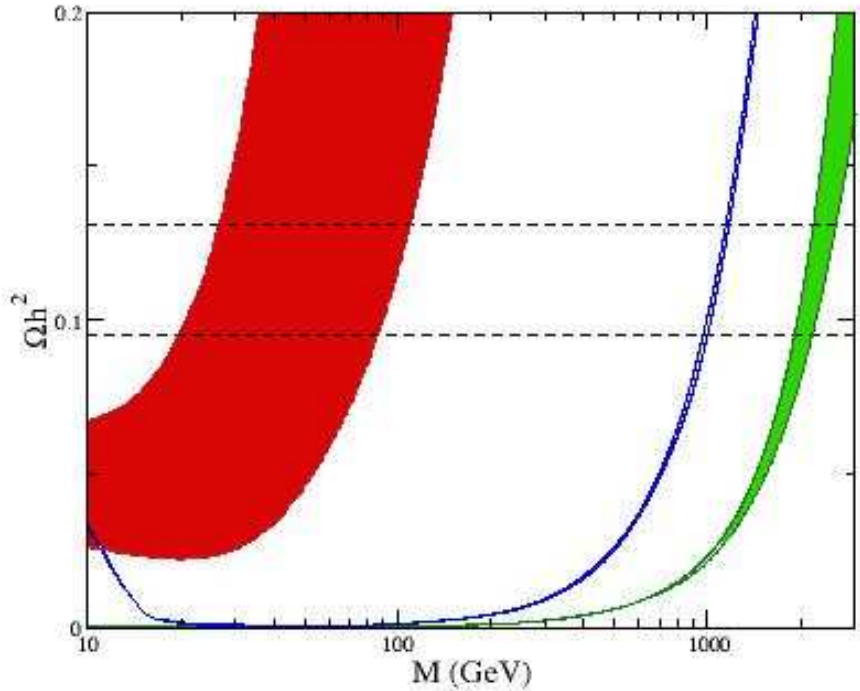}
\caption{The three bands show the contribution to $\Omega h^2$ from
pure Bino LSP with $0.3<M_1/m_{{\tilde e}_R}<0.9$ (red band), Higgsino LSP
with $1.5<m_{\tilde t}/\mu<\infty$ (blue band) and Wino LSP
with $1.5<m_{{\tilde \ell}_L}/M_2<\infty$ (green band).}
\label{fig1}
\end{figure}

In the limit of pure Higgsino, the dominant annihilation
channel is into gauge bosons. Since charged and neutral Higgsino states are
nearly mass-degenerate, co-annihilation is essential~\cite{higgs}.
When $\mu$ is larger than $M_W$, the effective annihilation cross
section and the Higgsino contribution to $\Omega$ are well approximated
by (see appendix A)
\beq
\langle \sigma_{eff} v\rangle
=\frac{g^4}{512 \pi \mu^2} \left(21+3 \tan^2 \theta_W+11\tan^4\theta_W\right),
\label{shig}
\eeq
\beq
\Omega_{\hino} h^2 =0.10 \left( \frac{\mu}{1 \tev}\right)^2 ,
\label{ohig}
\eeq
where in $\Omega_{\hino}$
we have suppressed a small logarithmic dependence on $\mu$.
The annihilation into $\bar tt$ pairs through stop exchange is less important,
as it is evident from
fig.~\ref{fig1}, where we show the Higgsino relic abundance: the narrow band
corresponds to a variation of the stop mass
from $m_{\tilde t}=1.5 \mu$ to infinity. In the mass region allowed by LEP
($\mu \gsim 100\gev$), we have the opposite problem of the Bino: the
relic abundance is too low, unless $\mu$ is about 1 TeV. Such large values
of $\mu$ require a significant fine-tuning in order to reproduce the weak scale
and are at odds with the original motivation of supersymmetry. They are
acceptable in the case of Split Supersymmetry~\cite{split1,split2,split3},
where supersymmetry is
not required to solve the hierarchy problem. Moreover, such large values of
the LSP mass typically
bring the supersymmetric spectrum beyond the reach of the LHC.

As in the case of the Bino, by
invoking solutions that depend on initial cosmological
conditions,
we can obtain a correct Higgsino abundance with
lower values of $\mu$, . Since now we want to increase the Higgsino abundance, rather than
depleting it, we have to assume that the super-weakly interacting relic
(gravitinos or other exotic particles)
is not the LSP, but it decays into Higgsinos.
The decay process has to occur after Higgsinos have decoupled, but
early enough not to upset the nucleosynthesis predictions. The final relic
abundance will of course depend on the initial gravitino density or,
ultimately, on $T_{RH}$.

 \subsection{Wino}

The Wino can be the LSP in anomaly mediation~\cite{anomaly,anomaly2}.
In the case of pure state, the
dominant annihilation is into gauge bosons, with a contribution from
fermion--antifermion channel through scalar exchange. Coannihilation among
the different states in the Wino weak triplet is important.
In the limit in which the Wino mass $M_2$ is larger than $M_W$,
the effective annihilation cross
section and the Wino contribution to $\Omega$ are well approximated
by (see appendix A)
\beq
\langle \sigma_{eff}v\rangle= \frac{3g^4}{16\pi M_2^2},
\label{swin}
\eeq
\beq
\Omega_{\wino} h^2 =0.13 \left( \frac{M_2}{2.5 \tev}\right)^2.
\label{owin}
\eeq
In $\Omega_{\wino}$
we have suppressed a small logarithmic dependence on $M_2$.
The
Wino relic abundance is shown in fig.~\ref{fig1},
where the narrow band corresponds to varying the left slepton masses from
$m_{{\tilde \ell}_L}=1.5 M_2$ to infinity, and taking the other supersymmetric
scalars very heavy.
The situation is analogous
to the Higgsino case, and actually it is even more extreme, because values
of $M_2$ as large as 2.5~TeV are required to reproduced the observed value
of the dark-matter density. In the case of anomaly mediation, the value
of the gravitino mass is adequate to generate an acceptable non-thermal
$\wino$ population to account for the dark matter, from gravitino or
moduli decay~\cite{rand,ghergh}.

\subsection{The Dark-Matter Impasse and the Well-Tempered Neutralino}

Figure~\ref{fig1} illustrates the meaning of the
supersymmetric dark-matter impasse. Before
LEP, values of $M_{1,2}$ and $\mu$ below $M_Z$ were allowed and supersymmetry
could explain the weak scale without much fine tuning. At that time,
a description of neutralinos in terms of pure states was not appropriate,
and the LSP was naturally a mixture.
Its relic abundance cannot be directly read from fig.~\ref{fig1}, because
more annihilation channels are possible for mixed states. However, it is
clear from fig.~\ref{fig1} that the correct value of $\Omega_{\rm DM}$ could
be reproduced, for mixed states and with light sleptons,
in a broad range of parameters. Certainly the prediction
for $\Omega_{\rm DM}$ could vary significantly, but the crucial point is
that this variation did not have a {\it critical} behaviour with the underlying
soft terms. On the other hand, the options left open after LEP, such as
$\tilde \tau$-coannihilation
or Higgs-resonance, give a prediction of $\Omega_{\rm DM}$
which is {\it critically} sensitive on some soft parameters.

The supersymmetric dark-matter impasse is very similar to the naturalness
problem. Before LEP, the prediction for $M_Z$ could certainly vary with
the parameters, and the physical value could be obtained only for
particular choices. However, the physical value of $M_Z$ did not imply a
special critical sensitivity on soft parameters. After LEP, the correct value
of $M_Z$ can still be reproduced, but the result has, at best, few-percent
sensitivity on parameter variations.
As the soft terms
are varied within the experimentally-allowed region,
in almost all cases one finds
$M_Z=0$ (no electroweak breaking) or $M_Z\gsim \tev$. The physical value
of $M_Z$ is obtained only for critical values of the soft terms, and
this requires a certain amount of fine tuning.

As we vary the values of the soft terms in the experimentally-allowed region
without accepting excessive fine tunings, we are
finding that $\Omega_{\rm DM}$ is typically too large (Bino LSP) or
too small (Higgsino or Wino LSP). Since $\Omega_{\rm DM}$ is a continuous
function of parameters, it is clear that, along the critical lines that
separates $\bino$/$\hino$ LSP and $\bino$/$\wino$
LSP, we can obtain the correct
value of the dark-matter density. This may appear as a fine-tuning but, if
supersymmetry is responsible for dark matter, LEP has forced us towards
such tuned regions. If we consider the general supersymmetric parameter
space, without bias in favour of universal/unified soft terms, these regions
appear as some of
the most plausible option for dark matter, still allowed after
LEP.

We will therefore consider in this paper
the case in which the LSP neutralino is mass degenerate or
maximally mixed with other states
(either $\bino$/$\hino$ or $\bino$/$\wino$), although the relevant
soft terms are typically larger than $M_Z$. This requires a precise
relation among parameters ($|M_1|\simeq |\mu|$ or $|M_1|\simeq |M_2|$)
and therefore
we will refer to this case as the ``well-tempered neutralino''.
We will start with a general classification, but many considerations
about mixed or degenerate heavy neutralinos have already been made in
the literature. The
case of the $\bino$/$\hino$ tempered neutralino was studied in the context
of focus-point~\cite{focusdm}, non-universal soft terms~\cite{nonunivdm},
and Split Supersymmetry~\cite{split2,pierce,mas}.
The case of $\bino$/$\wino$ tempered
neutralino was previously considered in refs.~\cite{mas,baer,Birkedal-Hansen:2001is},
assuming non-minimal gaugino mass relations.
We will present here
a class of theories where
the relation $|M_1|\simeq |M_2|$ can be obtained with conventional
gaugino-mass boundary conditions, like anomaly mediation or gaugino
unification.

\section{The Well-Tempered Bino/Higgsino}
\label{secbh}

In the case of the well-tempered $\bino /\hino$, the Wino can be decoupled
and the effective $3\times 3$ neutralino mass matrix in the basis
$\bino , \hino_1 \equiv(\hino_u -\hino_d )/\sqrt{2},\hino_2 \equiv
(\hino_u +\hino_d )/\sqrt{2}$ is
\begin{eqnarray}
&&{\cal M}=\pmatrix{ M_1 & -\frac{s_\beta +c_\beta}{\sqrt{2}}s_WM_Z&
 \frac{s_\beta -c_\beta}{\sqrt{2}}s_WM_Z \cr
-\frac{s_\beta +c_\beta}{\sqrt{2}}s_WM_Z &\mu &0\cr
\frac{s_\beta -c_\beta}{\sqrt{2}}s_WM_Z & 0 & -\mu }\nonumber\\
&&-\frac{M_W^2}{2M_2}\pmatrix{0&0&0\cr 0&1+s_{2\beta} & c_{2\beta}\cr
0& c_{2\beta}&1-s_{2\beta}}+{\cal O}\left( \frac{1}{M_2^2}\right)
.
\label{massm}
\end{eqnarray}
Here and in the following $s_\beta \equiv \sin\beta$,
$c_\beta \equiv \cos\beta$, $s_W \equiv \sin\theta_W$, etc.
The general discussion of the diagonalization of ${\cal M}$,
including the effect of the CP-violating phase, is
given in appendix B. For real $M_1$ and $\mu$, the well-tempered condition
can be achieved in two cases: $M_1 \simeq \mu$ and $M_1 \simeq -\mu$.
When $|M_1|$, $|\mu|$ and their difference are larger than $M_Z$, the
eigenvalues of ${\cal M}$, dropping terms suppressed by $M_1$,
are approximately given by
\beq
M_1 +\theta_\pm^2 \left( M_1 \mp \mu \right) ,~~~
\pm\mu -\theta_\pm^2 \left( M_1 \mp \mu \right) ,~~~ \mp \mu ,
\label{pix}
\eeq
\beq
\theta_\pm =\frac{(s_\beta \pm c_\beta ) s_W M_Z}{\sqrt{2} (\mu \mp M_1)},
\label{mix}
\eeq
where the $\pm$ sign refers to the cases $\mu \simeq \pm M_1$.
The orthogonal matrix $N$ that makes $N{\cal M}N^T$ diagonal is
\beq
N=\pmatrix{1-\frac{\theta_+^2}{2}-\frac{\theta_-^2}{2}&\theta_+ &\theta_- \cr
-\theta_+ & 1-\frac{\theta_+^2}{2} & -\theta_+ \theta_- \frac{(M_1+\mu)}{2\mu}
\cr -\theta_- & \theta_+ \theta_- \frac{(M_1-\mu)}{2\mu} &
1-\frac{\theta_-^2}{2}}.
\label{nrot}
\eeq
If $\mu \simeq  M_1$, the Bino and $\hino_1$ are mixed with an angle
$\theta_+$, but $\hino_2$ is nearly a pure state.
If $\mu \simeq  -M_1$, the mixing occurs between $\bino$ and
$\hino_2$ with mixing
angle $\theta_-$, while $\hino_1$ is nearly a pure state.
However notice
that if $\tan\beta =1$, $\hino_2$ becomes an exact mass eigenstate (to
all orders in $M_2$) which never mixes with $\bino$. Indeed,
$\theta_-$ vanishes when $\tan\beta =1$.
At large $\tan\beta$, the sign of $\mu$ is unphysical and the two
cases $\mu \simeq \pm M_1$ give identical results.

When $|\mu \pm M_1|<(s_\beta \mp c_\beta)s_W M_Z/\sqrt{2}$,
eqs.~(\ref{pix})--(\ref{mix}) are
no longer valid. In this case the $\bino$-$\hino_1$ mixing angle
(for $\mu$ and $M_1$ with same sign) or the $\bino$-$\hino_2$ mixing angle
(for $\mu$ and $M_1$ with opposite sign) become maximal.
The mass eigenvalues become
\beq
M_1+ \frac{\left( s_\beta \pm c_\beta \right)}{\sqrt{2}} s_W M_Z
+ \left( 1\mp s_{2\beta}\right) \frac{s_W^2 M_Z^2}{8M_1},
\label{cipo1}
\eeq
\beq
M_1- \frac{\left( s_\beta \pm c_\beta \right)}{\sqrt{2}} s_W M_Z
+ \left( 1\mp s_{2\beta}\right) \frac{s_W^2 M_Z^2}{8M_1},
\eeq
\beq
-M_1-\left( 1\mp s_{2\beta}\right) \frac{s_W^2 M_Z^2}{4M_1}.
\label{cipo2}
\eeq
Notice that two neutralino masses are split by terms ${\cal O}(M_Z)$.

\begin{figure}
\centering
\includegraphics[width=0.7\linewidth]{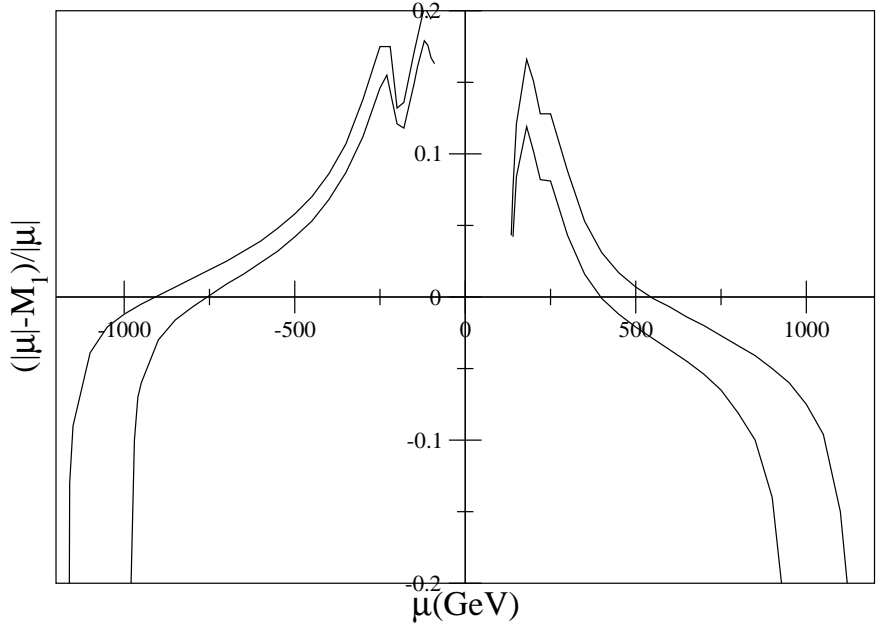}
\caption{The parameters of the well-tempered $\bino$/$\hino$ consistent
with the dark-matter constraint within 2$\sigma$. We have taken $\tan\beta=2$,
$m_H=115\gev$,
and heavy supersymmetric scalars, and chosen the convention $M_1>0$.
We have considered only $|\mu|>100\gev$, to satisfy the experimental
limit on chargino masses.}
\label{figbh}
\end{figure}

The relic abundance of the tempered $\bino$/$\wino$ is determined by
coannihilation among the 3 neutral states and one chargino with mass
approximately equal to $\mu$. Coannihilation of different species
$\chi_i$ is effective as long as the scattering processes
$\chi_i q \leftrightarrow \chi_j q^\prime$ (where $q$ and $q^\prime$
are relativistic particles) are in equilibrium. This requires
$\Gamma_S >H$, where $H$ is the Hubble constant today,
the scattering rate is $\Gamma_S \simeq
T^3\theta^2/M_\chi^2$ and $M_\chi$, $\theta$ are the typical $\chi$ mass
and mixing angle. Therefore our discussion is valid as long as
$\theta \gg (x_f \sqrt{g_*}M_\chi /M_{\rm Pl})^{1/2}$.

An analytic approximation of the relic abundance of well-tempered neutralinos
is described in appendix A, in the limit of small
mixing angle. The numerical result for
the parameters required to obtain the correct value of
$\Omega_{\rm DM} h^2$, using
the DARKSUSY package~\cite{darksusy},
is shown in fig.~\ref{figbh}. With about 10\%
degeneracy between $\mu$ and $M_1$, one can overcome the supersymmetric
dark-matter impasse and reproduce the correct relic
density for soft terms in the few-hundred GeV range.
We can also have
a dark-matter neutralino with $|\mu|<M_1$, and $\mu$ substantially
smaller than 1~TeV. This requires a large mixing
angle with the $\bino$ to reduce the Higgsino annihilation cross
section. This is why in  fig.~\ref{figbh}, the region with $|\mu|<M_1$
is more prominent in the positive-$\mu$ branch since, in the negative-$\mu$
branch, $\theta_-$ is suppressed by the moderate value of $\tan\beta$.
The dip of the negative-$\mu$ branch in fig.~\ref{figbh} corresponds
to the threshold for $t\bar t$ production ($m_{\chi^0}>m_t$), below
which a higher degree of degeneracy is needed to obtain the same value
of $\Omega_{\rm DM}$. For small $\mu$, the behaviour of the curves with
positive and negative $\mu$ is quite distict. Indeed, when
eqs.~(\ref{cipo1})--(\ref{cipo2}) are approximately valid, the mass
splitting in the positive-$\mu$ branch is $(t_\beta +1)/(t_\beta -1)$
times larger than in the negative branch. For $\tan\beta =2$ (the case shown
in fig.~\ref{figbh}) the mass splitting for moderate and positive $\mu$
becomes large enough to make coannihilation irrelevant. Then the dominant
annihilation channel is into Higgs and longitudinal gauge
bosons. This explains why the
$t\bar t$ threshold is less important for $\mu >0$, while the Higgs
threshold has a dramatic effect.

When $\mu$ is very close to its minimum value determined by the
limit on chargino masses, we cannot really talk about a well-tempered
neutralino, since $\mu$ and $M_1$ are comparable to $M_Z$. In this
case, the neutralino becomes a natural mixture of current eigenstates.
The supersymmetric dark-matter impasse can be resolved in this region,
where $\mu\sim M_1 \sim M_Z$. However, experimental bounds on chargino masses
strongly limit the size of this region.

The well-tempered $\bino$/$\hino$ is favourable for dark-matter
detection~\cite{mas}.
The spin-indepen-dent neutralino-proton
cross section mediated by the Higgs boson is given by~\cite{barb}
\beq
\sigma_p = \left( \frac{115 \gev}{m_H}\right)^4 \gamma^2 ~5.4 \times 10^{-43}
~{\rm cm}^2
\eeq
\beq
\gamma = \frac{t_W}{\sqrt{2}} N_{11} \left[
N_{12} (c_\beta +s_\beta)+N_{13} (c_\beta -s_\beta)\right] ,
\eeq
where $N$ is the matrix that diagonalizes $\cal{M}$. When
the mixing is moderate and eqs.~(\ref{pix})--(\ref{mix}) are valid, we find
\beq
\gamma \simeq t_W^2 M_W \frac{M_1+\mu s_{2\beta}}{\mu^2-M_1^2}
\eeq
At large $\tan\beta$, the detection rate is independent of $\tan\beta$ and
on the sign of $\mu$. For small $\tan\beta$, $\gamma \simeq
t_W^2 M_W /(\mu -M_1)$ is significant for same-sign $M_1$--$\mu$
($\bino$--$\hino_1$ mixing), but
it is suppressed for opposite sign ($\bino$--$\hino_2$ mixing).
On the other hand, when $|\mu \pm M_1|<(s_\beta \mp c_\beta)s_W M_Z/\sqrt{2}$
and we have
maximal mixing, then
$\gamma =t_W (c_\beta \pm s_\beta )/(2\sqrt{2})$, in the cases of
same-sign and opposite-sign, respectivey.

The well-tempered $\bino$/$\hino$ can also be detected through neutrino fluxes
from the sun~\cite{mas},
since its coupling to the $Z$ is non-vanishing, whenever the
mixing angles $\theta_\pm$ are large. It may also lead to observable
signals in antimatter cosmic rays from annihilation in the halo~\cite{mas}.
Of course,
the larger the coannihilation effect, the smaller the LSP annihilation cross
section, and therefore the more suppressed are the indirect dark-matter
signals from annihilation in the halo.

\section{The Well-Tempered Bino/Wino}

The well-tempered $\bino$/$\wino$ state is effectively described by the
$2\times 2$ mass matrix
\beq
{\cal M}=\pmatrix{M_1 & 0\cr 0 & M_2}-s_{2\beta} \frac{M_Z^2}{\mu}
\pmatrix{s_W^2 & -s_Wc_W\cr -s_Wc_W & c_W^2} +{\cal O}
\left( \frac{1}{\mu^2}\right) ,
\eeq
with $|M_1|\simeq |M_2|$.
The general discussion including CP-violating effects is given in
appendix B, while here we consider the case of real parameters. Notice that the
leading term in the $1/\mu$ expansion vanishes at large $\tan\beta$.
Therefore it is useful to give the expressions of the mass eigenvalues and the
$\bino$/$\wino$ mixing angle, including terms $1/\mu^2$. We define
\beq
\theta \equiv\frac{s_{2W} s_{2\beta}M_Z^2}{2\mu \Delta M_1} ,~~~~
\delta \equiv \frac{s_{2W}M_Z^2}{2\mu^2 \Delta},~~~~
\Delta \equiv \frac{M_2-M_1}{M_1}. \label{tett}
\eeq
First consider the case in which $M_2-M_1$ is not too small, such that
both $\theta$ and $\delta$ are smaller than one. Then
the $\bino$/$\wino$ mixing angle is
\beq
-N_{12}=\theta +\left( \frac{1}{t_W}-t_W \right) \theta^2
+\delta ,
\eeq
where $N$ is the orthogonal matrix that diagonalizes the neutralino mass
matrix.
Notice that $\theta$ approximately describes the mixing angle for small
or moderate $\tan\beta$. For large $\tan\beta$, the mixing angle
appears only at  ${\cal O}(1/\mu^2)$ and it is given by $\delta$.
For $M_1\simeq M_2$ the mixing angle can become large, while for
$M_1\simeq -M_2$ the mixing angle is negligible.
The mass eigenvalues of the neutral and charged states are
\bea
&m_{\chi_1}=M_1 \left[ 1-\Delta \left( t_W \theta +\theta^2 +t_W \delta \right)
\right] \cr
&m_{\chi_2}=M_1 \left[ 1+\Delta \left( 1-\frac{\theta}{t_W} +\theta^2
-\frac{\delta}{t_W} \right)
\right] \cr
&m_{\chi^+}=M_1 \left[ 1+\Delta \left( 1-\frac{\theta}{t_W}
-\frac{\delta}{t_W} \right)
\right] .
\eea
Notice that $m_{\chi^+}$ and $m_{\chi_2}$ are split only by
$1/\mu^2$ effects, and therefore the two states are nearly mass degenerate.

When $M_1-M_2$ is so small that $\theta$ or $\delta$ become larger than
unity, the photino is a mass eigenstate and the
$\bino$/$\wino$ mixing angle is approximately equal to $\theta_W$.
The mass eigenvalues are
\bea
&m_{\chi_1}=M_1 \cr
&m_{\chi_2}=M_1\left(1-\frac{s_{2\beta}M_Z^2}{\mu M_1} -\frac{M_Z^2}{\mu^2}
\right) \cr
&m_{\chi^+}=M_1\left(1-\frac{s_{2\beta}M_W^2}{\mu M_1} +\frac{M_W^2}{\mu^2}
\right) .
\eea
In this case, all states are split at ${\cal O}(1/\mu)$, unless $\tan\beta$
is very large.

\begin{figure}
\centering
\includegraphics[width=0.7\linewidth]{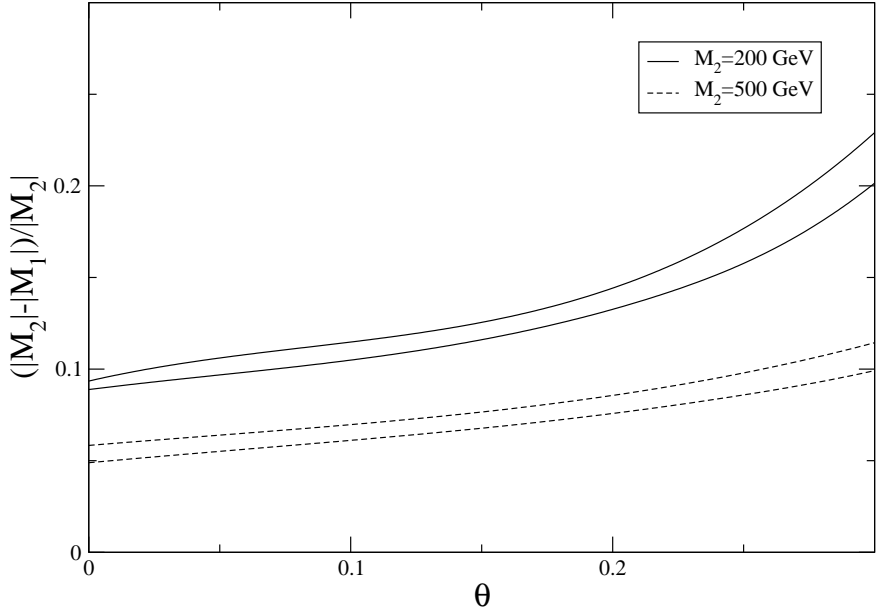}
\caption{The bands show the
parameters of the well-tempered $\bino$/$\wino$ consistent
with the dark-matter constraint within 2$\sigma$. We have taken
heavy supersymmetric scalars.}
\label{figbw}
\end{figure}

The relic abundance of the well-tempered $\bino$/$\wino$ is
determined in terms of the two masses $M_{1,2}$ and the mixing angle.
An analytic approximation is presented in appendix A, in both cases
of non-vanishing and vanishing mixing angle.
The numerical
result, using DARKSUSY~\cite{darksusy}, is
shown in fig.~\ref{figbw}. For a degeneracy
between $M_1$ and $M_2$ of about 10{\%}, dark-matter neutralinos can be brought
within the mass range of few hundreds GeV. For larger mixing angles,
the degree of degeneracy can be reduced. Sizable mixing angles cannot be
obtained in the case in which $M_1$ and $M_2$ have opposite sign, but they
occur for generic phases between $M_1$ and $M_2$, as discussed in appendix B.

\subsection{Connection with High-Energy Soft Terms}

The relation $|M_1|\simeq |M_2|$, necessary for well-tempered $\bino$/$\wino$,
has to hold at the weak scale. At first sight, this appears inconsistent
with either gaugino mass unification or anomaly mediation. However, this
is not the case if $\mu$ is much larger than $M_{1,2}$. Indeed, renormalization
effects below the scale of the heavy Higgs doublet mass ($m_A$) generate
additive
contributions to the gaugino masses proportional to $\mu$. Let us consider
the one-loop corrected gaugino masses, including terms enhanced by a
logarithm or proportional to $\mu$~\cite{bagger}
\bea
M_1=& M_1(m_A^2)\left[ 1+\frac{\alpha}{8\pi c_W^2}
\left( 11 \log \frac{{\tilde m}_q^2}{m_A^2}+9\log \frac{{\tilde m}_\ell^2}
{m_A^2} +\log\frac{\mu^2}{m_A^2}\right) \right]
\nonumber \\
&+\frac{\alpha}{8\pi c_W^2} \mu s_{2\beta} f\left( \frac{\mu^2}
{m_A^2}\right)
\label{m1loop}
\eea
\bea
M_2=& M_2(m_A^2)\left[ 1+\frac{\alpha}{8\pi s_W^2}
\left( 9 \log \frac{{\tilde m}_q^2}{m_A^2}+3\log \frac{{\tilde m}_\ell^2}
{m_A^2} +\log\frac{\mu^2}{m_A^2}-12\log\frac{M_2^2}{m_A^2} \right)
\right]
\nonumber \\
&+\frac{\alpha}{8\pi s_W^2} \mu s_{2\beta} f\left( \frac{\mu^2}
{m_A^2}\right)
\label{m2loop}
\eea
\beq
f(x)=\frac{2\log x}{1-x},
\eeq
where all
coefficients on the right-hand sides of eqs.~(\ref{m1loop})--(\ref{m2loop})
have to be evaluated at the scale $m_A$. The gaugino masses at the scale $m_A$
are obtained by evolving high-energy boundary conditions using
renormalization-group flow. For gaugino-mass unification and for
anomaly mediation, respectively, we have
\bea
M_1(m_A)=\frac{5\alpha (m_A)}{3 c_W^2 \alpha_{\rm GUT}} M_G,
~M_2(m_A)=\frac{\alpha (m_A)}{s_W^2 \alpha_{\rm GUT}} M_G
~&{\rm (gaugino~unif.)} \\
M_1(m_A)=\frac{11\alpha (m_A)}{4\pi c_W^2}m_{3/2},
~M_2(m_A)=\frac{\alpha (m_A)}{4\pi s_W^2}m_{3/2}
~&{\rm (anomaly~med.)}
\eea
where $M_G$ is the common gaugino mass at the unification scale and
$m_{3/2}$ is the gravitino mass.

In our numerical analysis, we have actually resummed the logarithms
using the renormali-zation-group equations for Split Supersymmetry,
but the one-loop results in
eqs.~(\ref{m1loop})--(\ref{m2loop})
are useful for our discussion because they
explicitly exhibit an essential feature.
The quantum corrections proportional to $\mu$ can significantly modify
the high-energy gaugino-mass boundary condition (for $\mu \gg M_{1,2}$) and
they can realize the well-tempered mass relation.
In the case of gaugino-mass unification, the relation $M_1=M_2$ is
obtained for $\mu s_{2\beta} /M_G=49$ (for $m_A/\mu=10$) or 25
(for $m_A/\mu=100$); the relation $M_1=-M_2$ is
obtained for $\mu s_{2\beta} /M_G=81$ (for $m_A/\mu=10$) or 40
(for $m_A/\mu=100$).
This requires a certain hierarchy between soft terms, for which we have
no good theoretical justification.

\begin{figure}
\centering
\includegraphics[width=0.7\linewidth]{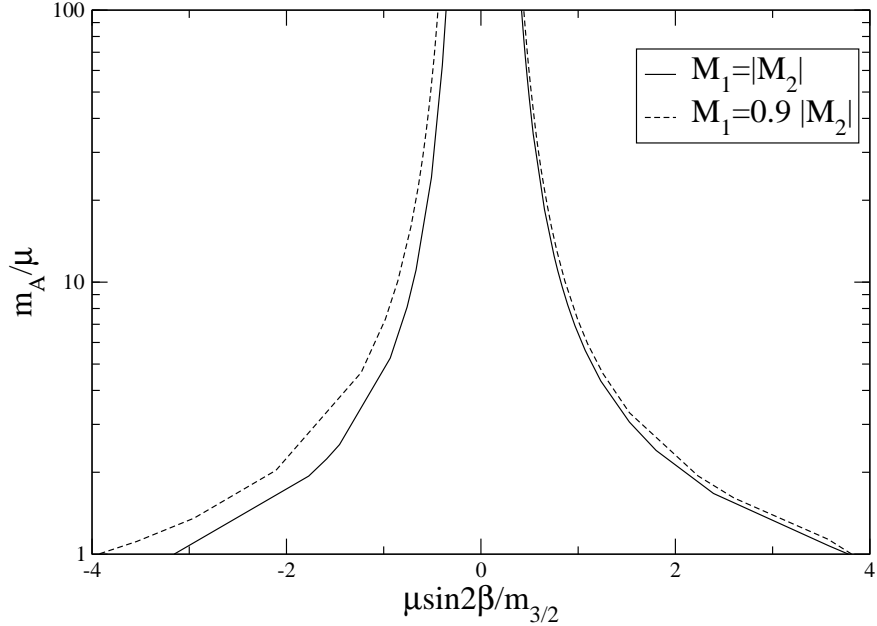}
\caption{The relation between $\mu$, $m_A$ and $m_{3/2}$ necessary to obtain
a well-tempered $\bino$/$\wino$ with $0<(|M_2|-|M_1|)/|M_2|<0.1$, assuming
gaugino masses from anomaly mediation.}
\label{figam}
\end{figure}

More interesting is the
case of anomaly mediation because, as shown in fig.~\ref{figam}, the
$\bino$/$\wino$
well-tempered relation is obtained for values of $\mu$ of the order of
the gravitino mass $m_{3/2}$. As we will show in the next section, this
can be naturally obtained in simple models of Split Supersymmetry.
The two branches of solutions shown in fig.~\ref{figam} correspond
to the case in which $M_1$ and $M_2$ have equal sign (negative $\mu$)
and opposite sign (positive $\mu$).

\section{The Simplest Model of Split Supersymmetry}
\label{sec4}

Here we want to discuss one of
the simplest construction of Split Supersymmetry and also show that it
can lead to a well-tempered $\bino$/$\wino$. Let us consider the case
in which supersymmetry is broken by
an auxiliary field in the combination
$F^\dagger F =M_S^4 \theta^2 {\bar \theta}^2$,
generating through direct mediation
soft masses for dimension-2
operators: squark and slepton square masses and the $B_\mu$ parameter, of
the order of ${\tilde m^2}= M_S^4/M^2$.
We are interested in a mediation scale $M$ in the range
$M_{\rm GUT}<M<M_{\rm Pl}$,
where $M_{\rm Pl}$ is the reduced Planck mass,
so that the new dynamics cannot affect
gauge-coupling unification. If the supersymmetry-breaking sector
contains no singlets and the breaking is mainly along the $D$-component
$F^\dagger F$, the direct mediation of the dimension-3 soft-term
operators (gaugino masses, $A$ and $\mu$ terms) is suppressed~\cite{split3}.
Anomaly
mediation then gives the leading contribution. The gaugino masses
are generated by the coupling of the chiral compensator $\Phi$ to
the gauge kinetic term and are
given by $M_g (Q)=(d \log g /d \log Q)m_{3/2}$~\cite{anomaly,anomaly2}.

There are two couplings of the chiral compensator to the Higgs superfields
that can generate a $\mu$ term:
\beq
{\cal L}=\lambda \int d^4\theta \Phi^\dagger \Phi H_u H_d +
\left( \rho \int d^2\theta \Phi^3 H_u H_d +{\rm h.c.}\right) .
\eeq
Performing a field rescaling $\Phi H_{u,d}\to H_{u,d}$, the Lagrangian
becomes
\beq
{\cal L}=\lambda \int d^4\theta \frac{\Phi^\dagger}{ \Phi} H_u H_d +
\left( \rho \int d^2\theta \Phi H_u H_d +{\rm h.c.}\right) .
\eeq
Since $\Phi=1+\theta^2 m_{3/2}$, we obtain the following contributions
to the $\mu$ and $B_\mu$ terms, respectively,
\beq
\mu =\lambda m_{3/2} +\rho , ~~~B_\mu  =\lambda m^2_{3/2} -\rho m_{3/2}.
\label{strit}
\eeq
The parameter $\rho$ is dimensionful and it is a manifestation
of the $\mu$ problem,
since it originates from  supersymmetric-invariant dynamics and its value
is not, in principle, related to $m_{3/2}$. On the other hand, the
coupling $\lambda$ corresponds to the natural solution of the $\mu$-problem
in supergravity~\cite{giumas}.

If $\rho$ vanishes and the contribution to $B_\mu$
from direct coupling to the
supersymmetry-breaking sector is suppressed, then from \eq{strit}
we find the relation
$B_\mu =\mu m_{3/2}$. Minimization of the Higgs potential gives
$s_{2 \beta} =2B_\mu /m_A^2$, where $m_A$
is the mass of the heavy Higgs doublet, which is of the order of $\tilde m$.
Therefore, in this case, an acceptable value of
$\tan\beta$ requires ${\tilde m}\sim m_{3/2}$ and the scale of
supersymmetry-breaking mediation $M$ is near the Planck scale. Also notice that
in this case we find
\beq
\frac{\mu s_{2 \beta}}{m_{3/2}} =\frac{2\mu^2}{m_A^2}.
\label{relg}
\eeq
This relation corresponds to the model considered in
ref.~\cite{anomaly2}.
It is compatible with a well-tempered
$\bino$/$\wino$ (with opposite signs for $M_1$ and $M_2$)
when $m_A= 0.57 \mu $.
Through \eq{relg} this predicts $\mu s_{2\beta} = 6.1 m_{3/2}$.
Also notice that, if $\rho$
were the dominant source for $B_\mu$,
we would obtain
$B_\mu =-\mu m_{3/2}$, which leads to the relation $\mu s_{2 \beta} /
m_{3/2} =-2\mu^2/m_A^2$. This is compatible with a well-tempered
$\bino$/$\wino$ (with equal signs for $M_1$ and $M_2$)
when $m_A= 0.55 |\mu | $, leading to $\mu s_{2\beta}
= -6.6 m_{3/2}$.
In general, however, the direct mediation contribution to $B_\mu$ is
significant and the relation in \eq{relg} will not be strictly exact.

To summarize, the Simplest Model of Split Supersymmetry (SMS) is
characterized by: {\it (i)} Squark, slepton masses and $B_\mu$ of
typical size $\mtil = k m_{3/2}$, with $1\lsim k \lsim M_{\rm
Pl}/M_{\rm GUT}$;
 {\it (ii)} $\mu$ parameter of typical size $\mu\sim m_{3/2}$;
{\it (iii)} Gaugino masses given by the anomaly-mediation relation
($M_{\tilde g} \sim m_{3/2} ~g^2/16\pi^2$), with large
corrections to $M_{1,2}$ proportional to $\mu s_{2\beta}$
given by eqs.~(\ref{m1loop})--(\ref{m2loop}).
Therefore, the SMS is described by 4 parameters ($\mtil$, $m_{3/2}$, $\mu$,
$\tan\beta$), with typical mass ratios, as specified above. It is, in our
view, the most straightforward construction of a model with Split
Supersymmetry. It is a variation of the model first considered in
ref.~\cite{anomaly2}, and then revived in ref.~\cite{wells}.

There is another limit, considered in ref.~\cite{split1}, where the
direct coupling of the Higgses to the supersymmetry-breaking
sector is not
suppressed, so that $B_\mu \epsilon \sim \tilde{m}^2$ where
$\epsilon$ is a parameter characterizing the size of PQ breaking.
This leads to $\tan
\beta \sim 1/\epsilon$, and also $\mu \sim \epsilon m_{3/2} \sim
m_{3/2}/\tan \beta$. In this case it is possible to have Higgsinos
near the TeV scale with large $\tan \beta$.

The SMS is minimal in the sense that essentially {\it any}
model of supersymmetry breaking, as long as it does not have singlets to
directly mediate gaugino masses, will produce this spectrum
without additional special structure to ``sequester" the scalar
masses from supersymmetry breaking. There are exceptions where, for good
physical reasons, the anomaly mediated contributions are
suppressed as described in ref.~\cite{split1} and in detail in
ref.~\cite{split3}. If supersymmetry is broken already in global limit,
$F_\Phi
= m_{3/2}$ and the anomaly mediated contribution is there, while
if supersymmetry is restored as gravity is decoupled, $F_\phi \to 0$ and
anomaly mediation is shut off. But there are certainly wide
classes of models of both types. SMS makes the scalars heavy
enough to eliminate all the usual problems of low-energy supersymmetry.

The requirement of an appropriate thermal dark-matter density can
be satisfied when the LSP Wino has a mass of about 2.5 TeV, or
with light Higgsinos near 1 TeV \footnote{Note that with the
gravitino near 100 TeV, the decay of the gravitino can dilute the
thermal relic abundance, so we must assume a low enough reheating
temperature for this not to happen. An alternative cosmological
history in this model is examined in \cite{Kaplan}.}. In either of
these cases, no observable signals can be expected at the LHC.
More appealing is the case in which the SMS leads to a
well-tempered $\bino$/$\wino$. The necessary relation among $\mu$,
$\mtil$ ($=m_A$), $m_{3/2}$, and $\tan\beta$ is illustrated in
fig.~\ref{figam}. It is interesting to notice that the SMS
expectations $\mu /m_{3/2}={\cal O}(1)$ and $\mtil /\mu ={\cal
O}(k)$ are exactly what is needed to obtain a well-tempered
$\bino$/$\wino$, whenever $s_{2\beta}$ is not too small ({\it
i.e.} $\tan\beta$ is not too large). Notice that for complex soft
parameters, the values of $|\mu| s_{2\beta}/m_{3/2}$ that lead to
a well-tempered $\bino$/$\wino$ vary, as we change the relative
phase $\varphi$ between $\mu$ and $m_{3/2}$, in the interval
$|\mu_+|\le |\mu| \le |\mu_-|$, where $\mu_\pm$ are the solutions
in the positive and negative branches shown in fig.~\ref{figam}.

Once the well-tempered condition is verified, and the parameters satisfy
the relation $|M_1|=|M_2|$, then the ratio between the gaugino
and gravitino mass is fixed, and it is independent of the value of $\mu$,
up to small logarithmic effects, depending only on the phase
$\varphi$. We find $m_{3/2}/|M_2| =a_1(\varphi )$,
where $a_1( \pi)
\simeq 80$
(negative-$\mu$ branch) and  $a_1(0) \simeq 150$ (positive-$\mu$
branch), and intermediate values between 80 and 150 are obtained
with a generic phase $\varphi$.

The well-tempered condition eliminates one parameter, and therefore the
SMS can be described in terms of: {\it (i)} an overall mass scale, which
we take
to be the gaugino mass $|M_2|$; {\it (ii)}
the mass ratio $\mtil /M_2$, which describes the degree of mass
hierarchy or, in other words, the amount of Split Supersymmetry;
{\it (iii)} $\tan\beta$, which cannot be too large, or else
the radiative contribution
to gaugino masses proportional to $\mu$ is suppressed;
{\it (iv)}
the relative sign between $\mu$ and $m_{3/2}$ (or more generally,
for complex terms, the phase $\varphi$). The theory also selects an
allowed interval for the ratio $\mtil /|M_2|$:
\beq
\frac{a_2(\varphi )}{s_{2\beta}} <\frac{\mtil}{|M_2|}
< a_1(\varphi ) \frac{M_{\rm Pl}}{M_{\rm GUT}},
\label{flop}
\eeq
where $a_2( \pi) \simeq 260$ (negative $\mu$) and
$a_2( 0) \simeq 570$ (positive $\mu$).

The lower bound on  $\mtil /|M_2|$ is obtained
from the requirement that
the well-tempered condition can be achieved. The contribution
to gaugino masses $\mu/\mtil ~f(\mu^2/\mtil^2)$
in eqs.~(\ref{m1loop}) and (\ref{m2loop}) has a minimum for
$\mu =\mtil$, which determines the lower limit in \eq{flop}.
The upper bound on
$\mtil /|M_2|$ is obtained from the ratio $m_{3/2}/|M_2|$, using
$\mtil /|M_2| =a_1(\varphi ) k$ and $k \lsim M_{\rm Pl}/M_{\rm GUT}$.
Therefore, from \eq{flop} we infer that the SMS can vary from a mild to
a more accentuated version of Split Supersymmetry, with a mass hierarchy
between
supersymmetric scalars and fermions which is, however, never extreme.

\begin{figure}
\centering
\includegraphics[width=0.7\linewidth]{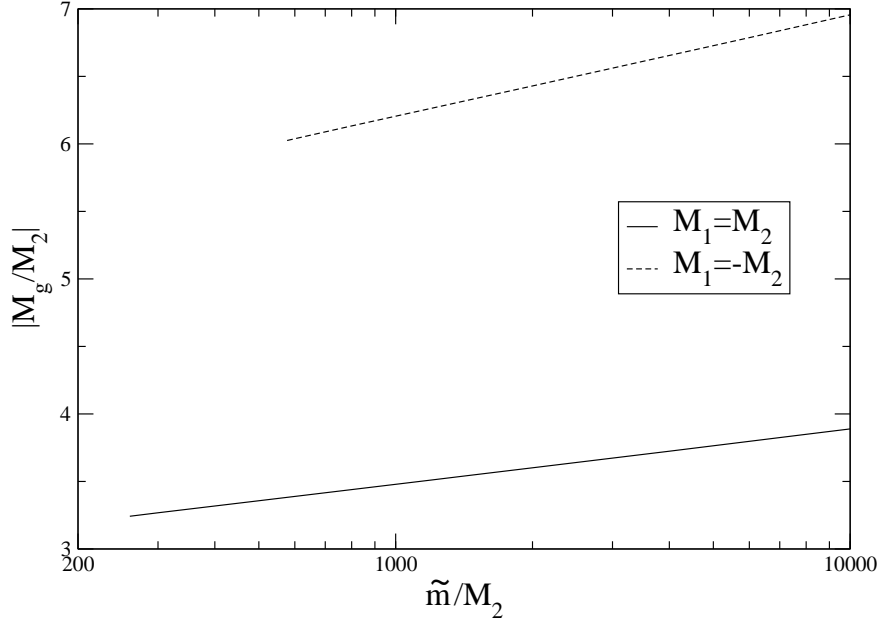}
\caption{The ratio $M_{\tilde g}/|M_2|$ between the gluino pole mass
and the gaugino mass as a function of $\mtil /|M_2|$ for
$\tan\beta =1$, assuming the
well-tempered $\bino$/$\wino$ relation in the negative-$\mu$ branch ($M_1=M_2$)
and the positive-$\mu$ branch ($M_1=-M_2$). The curves start from the
minimum value of $\mtil /|M_2|$ determined by \eq{flop}.}
\label{figglu}
\end{figure}

The mass of the gluino in the SMS
is also fixed in terms of the other parameters and
in fig.~\ref{figglu} we show the ratio between the gluino pole mass and the
gaugino mass as a function of $\mtil$. In the negative-$\mu$ branch, the ratio
$M_{\tilde g}/M_2$ is sufficiently small to allow
for a large mass range where both
the gluino and neutralinos/charginos are within the kinematic
reach of LHC. The situation
is less favourable in the positive-$\mu$ branch, where $M_{\tilde g}/M_2$
is almost twice as large. Nevertheless, this ratio is somehow smaller
than in the case of pure anomaly mediation, where
$M_{\tilde g}/M_2=7.1$~\cite{ghergh}.

The gluino lifetime is~\cite{slavwell}
\beq
\tau_{\tilde g} =\left( \frac{\mtil}{10^5\gev}\right)^4
\left( \frac{{\rm TeV}}{M_{\tilde g}}\right)^5 ~7\times 10^{-16}~{\rm sec}.
\label{lifeglu}
\eeq
Before decaying, the gluino will travel a mean distance
$d\sim \tau_{\tilde g} ~p/M_{\tilde g}$, where $p$ is the gluino momentum.
Therefore, from \eq{lifeglu}, we infer that the SMS gluino always
decays inside the detector since, as previously discussed, $\mtil /M_2$
is expected to lie between $10^2$ and $10^4$.
A vertex
displacement can be observed by LHC detectors if $\tau_{\tilde g}\gsim
10^{-12}$~sec, which implies
\beq
\mtil \gsim  \left( \frac{M_{\tilde g}}{{\rm TeV}}\right)^{5/4}~
600\tev .
\label{displ}
\eeq
Therefore, in a range of SMS parameters, measurements at the LHC
can determine the value of $\mtil$. However, it will be more difficult to
distinguish the SMS with $\mtil$ below the value in \eq{displ}
from a more conventional supersymmetry with scalars barely escaping
searches at the LHC. One possibility to disantangle the two scenarios
is offered by the search for
gluinonium~\cite{gluin}. In the limit of heavy scalars, the gluino decay
process is slower than the gluino annihilation inside the gluinonium
bound state. Particularly interesting is the $^3 S_1 ({\bf 8}_A)$
gluinonium state,
which cannot decay into two gluons (a final state with large QCD background
from dijets), but only into quark-antiquark pairs. Bottom and/or top
identification can help reducing the background. The $^3 S_1 ({\bf 8}_A)$
decay width into massless quarks is approximately~\cite{gluin}
\beq
\Gamma =\frac{81}{32}\alpha_s(M_{\tilde g})^5 M_{\tilde g} ,
\label{life}
\eeq
with about $1/6$ branching ratio in each quark channel.
If $\mtil$ is few times bigger than $M_{\tilde g}$, then $\Gamma$ is larger
than $\tau_{\tilde g}^{-1}$, and the gluinonium annihilates before its
gluino components can
decay. Observation of gluinonium could provide an indication
to discriminate between the SMS and more conventional supersymmetry, with
squarks relatively close in mass to the gluino.

The search for neutralinos and charginos at the LHC will be challenging.
These particles will be produced by Drell-Yan processes or via gluino
decay. Signals are very sensitive on the mass difference between the
states~\cite{baer} and they can be hard to detect when the decay products
become soft. More studies are needed on the detection of quasi-degenerate
neutralinos and charginos.

For complex soft parameters, the SMS will be well tested by present and
future experimental searches for electric dipole moments~\cite{edms}.
Notice that the low-energy gaugino masses $M_{1,2}$ are given by the sum of
two comparable contributions: one proportional to $m_{3/2}$ and one to
$\mu$, see eqs.~(\ref{m1loop})--(\ref{m2loop}), with a relative
phase $\varphi$. Therefore, in contrast to the studied versions of
Split Supersymmetry, the two physical CP-violating phases
${\rm arg} (M_{1,2}\mu)$ are different, and can be suppressed
if the $\mu$-contribution to gaugino masses is dominant. For $M_1$, the pure
anomaly-mediation contribution is equal to 0.8 (1.5) times the total value
of $M_1$ in the negative (positive) $\mu$-branch, and this will not
suppress ${\rm arg} (M_{1}\mu)$. However, we expect a certain suppression of
${\rm arg} (M_2\mu)$, since the pure
anomaly-mediation contribution to $M_2$ is equal to 0.3 ($-0.5$)
times the total, in the negative (positive) $\mu$-branch.
The SMS contribution to
electric dipole moments is smaller than in Split Supersymmetry, because
Higgsinos are heavier, but it can still be detected in future experiments.
If $\tilde m$ is not much larger than $m_{3/2}$, then the leading contribution to electric dipole moments in
the SMS comes from one-loop diagrams mediated by scalar particles. For large phases, the effect can be
detected in future experiments.

\section{Conclusions}

We have argued that, after the unsuccessful searches at LEP, it is no
longer justified to claim that the correct value of $\Omega_{\rm DM}$ is
a {\it natural} prediction of low-energy supersymmetry. Once the
experimental constraints are satisfied, low-energy supersymmetry typically
leads to a too large ($\bino$ LSP) or too small ($\hino$ or $\wino$ LSP)
value of $\Omega_{\rm DM}$. Notwithstanding, the minimal supersymmetric
model can successfully reproduce the correct value of $\Omega_{\rm DM}$.
However, apart from a small region with sleptons or with $\mu$ and $M_1$
very close to their present experimental limits, the prediction for
$\Omega_{\rm DM}$
requires a {\it critical} choice of parameters. It is critical in the
sense that it is highly sensitive on small parameter variations.
In particular, when we rely on coannihilation, $\Omega_{\rm DM}$
exponentially depends on the difference between two {\it a priori}
unrelated soft parameters. Of course this is not a problem of
consistency, but only a conceptual difficulty completely analogous to
the fine-tuning problem of electroweak-symmetry
breaking. In low-energy supersymmetry, the correct value of $M_Z$ can be
reproduced by the soft terms, but only at the price of critical
choices of the parameters ({\it i.e.} with few-percent sensitivity
on small variations). We view it
as disturbing that two of the most important conceptual achievements of
low-energy supersymmetry -- electroweak breaking and dark matter -- both
require special conspiracies on the values of the underlying soft parameters.
Of course, from a more optimistic point of view, one can view both problems
as the result of a fortuitous and unlucky occurrence that has determined values
of the soft terms highly untypical of the acceptable parameter space.
After all, it is admittedly remarkable and non-trivial that both values of
$M_Z$ and $\Omega_{\rm DM}$ can be successfully computed in terms of
soft-breaking parameters.

If we insist that the dark matter originates from a supersymmetric thermal
relic, free from any dependence on the cosmological evolution above the
freeze-out temperature $T_f$, then we have to accept that the soft terms
approximately satisfy some critical condition. We have argued that the
case of the {\it well-tempered neutralino} is one of the most plausible
of these choices. A well-tempered neutralino corresponds to the boundary
between a $\bino$ LSP and a $\hino$ or $\wino$ LSP, where $\Omega_{\rm DM}$
rapidly varies from being too large to being too small. We have studied
the properties of well-tempered neutralinos, also considering the case of
complex soft terms. Since in the general parameter space of low-energy
supersymmetry, well-tempered neutralinos are some of the most favourable
candidates for dark matter, we believe that there are good motivations for
dedicated studies on the detection of quasi-degenerate neutralinos and
charginos at the LHC.

We have also shown that the $\bino$/$\wino$ well-tempered condition
$|M_1|\simeq |M_2|$ can be satisfied in models with conventional
high-energy boundary conditions for the gaugino mass, taking a large
$\mu$ term. One of the most interesting models of this class is the
SMS, discussed in sect.~\ref{sec4}, which is the simplest realization
of Split Supersymmetry. In the SMS, direct couplings to a hidden sector
generate the supersymmetric scalar masses, and anomaly mediation generates
the masses for the supersymmetric fermions. The SMS is consistent with
a well-tempered $\bino$/$\wino$ and, once the appropriate relation is imposed,
we find specific predictions for the mass ratios. It is not an extreme version
of Split Supersymmetry, since the ratio between the common scalar masses
and the gaugino mass $\mtil /M_2$ varies between few $10^2$ and about $10^4$,
see \eq{flop}.
As a
function of $\mtil /M_2$, $\tan\beta$ and (for complex parameters)
the phase of $\mu$, the mass ratios $\mu /M_2$ and
$M_{\tilde g}/M_2$ are determined, see figs.~\ref{figam} and \ref{figglu}.

Finally, we want to remark that the supersymmetric dark-matter
impasse, discussed in sect.~\ref{sec1}, does not immediately apply
to Split Supersymmetry, since values of $\mu$ of about 1~TeV or
$M_2$ of about 2.5~TeV are perfectly acceptable, once we abandon
the naturalness criterion. Why then should we expect to have an extra tuning
to get well-tempered neutralinos? It is difficult to answer this question
without having a more precise notion of what the physical measure of
tuning actually is, but we can at least identify a competition between two factors.
If we scale up the Wino to 2.5~ TeV as the LSP, so there is no tuning for 
dark matter,
we are making the scalars heavier too, which makes electroweak
breaking more tuned. If we leave Winos in the hundreds of GeV range,
the scalars are lighter and electroweak breaking is 
less tuned but there is more tuning to get the dark matter.
At any rate, a 2.5 ~TeV Wino make Split
Supersymmetry invisible at the LHC (for conventional gaugino mass
relations). Therefore the collider detectability of Split
Supersymmetry relies on the existence of well-tempered neutralinos.

Of course this conclusion depends on our assumption that the dark
matter is entirely composed of the thermal relic. It is possible
that the cosmological history is more involved. For instance,
in the minimal version of Split Supersymmetry with gravitinos near $\sim
100 \tev$, for a high reheating temperature the gravitinos are
produced with sufficient abundance so that their (safe) decay
before nucleosynthesis nonetheless significantly dilutes the LSP abundance.
Depending on the presence of other light moduli fields with masses
around $\tilde{m}$, the LSPs can be repopulated, and so lighter
LSPs can form the dark matter, with a visible gluino at the LHC
\cite{Kaplan}. Alternately, there may be more than one source of
dark matter, for instance not just our LSP but also axions. It is
further plausible that the dark-matter abundance is ``environmentally"
selected for \cite{DMenv}. In this case, both the initial
amplitude of the axion as well as the scale of gaugino masses need
to be lowered in order to get a small enough dark-matter abundance, but it
is reasonable to expect comparable abundances from each one. This
would again mean that we could have a lighter LSP, now providing
an ${\cal O}(1)$ fraction of the DM, but again with a spectrum
visible at the LHC.

However, the experimental observation of Split Supersymmetry with a
well-tempered neutralino for dark matter would be very striking.
Even in the minimal version of Split Supersymmetry with anomaly mediated
gaugino masses and scalars in the range between $10^2 - 10^3$ TeV,
there would be evidence for an enormous tuning for electroweak
symmetry breaking of about 1 part in $\sim 10^6$. This would
already represent a severe blow against naturalness, only further
augmented by the additional tuning required for getting the
correct dark-matter abundance. Fortunately, the LHC will soon
begin to tell us what path Nature has chosen -- and whether the weak
scale will represent the triumphant return or final downfall of
naturalness.

\section*{Appendix A}

In this appendix we give the analytic expressions for annihilation
and coannihilation processes relevant to the cases of pure states and
well-tempered neutralinos, in the limit in which $M_W$ is much
smaller than the gaugino and Higgsino masses.

The relic abundance is given by
\beq
\Omega h^2 =\frac{688 \pi^{5/2} T_\gamma^3 (n+1) x_f^{n+1}}
{99\sqrt{5g_*} (H_0/h)^2 M_{\rm Pl}^3\sigma}=
\frac{8.7\times 10^{-11} \gev^{-2} (n+1) x_f^{n+1}}{\sqrt{g_*}\sigma},
\label{omsi}
\eeq
where $g_*$ is the number of degrees of freedom at freeze-out,
$H_0$ is the present Hubble constant and $T_\gamma$ is the CMB temperature.
Here $\sigma$ is related to the
thermal-averaged non-relativistic annihilation cross section by
\beq
\langle \sigma v\rangle =\sigma x^{-n},~~~~x=\frac{m_\chi}{T},
\eeq
and the freeze-out temperature $T_f$ ($x_f={m_\chi}/T_f$) is
\beq
x_f=X-\left( n +\frac{1}{2}\right) \log X, ~~~
X=25 +\log \left[ (n+1)\frac{g}{\sqrt{g_*}}m_\chi \sigma ~6.4 \times 10^6\gev
\right] ,
\eeq
where $g=2$ are the neutralino degrees of freedom and $m_\chi$ is its mass.

When $N$ states with mass $m_i$ ($m_1$ being the lightest)
 and equal number of degrees of freedom
participate in the coannihilation process, one defines~\cite{coann}
an effective cross section $\sigma_{eff}$ in terms of
the thermal-averaged
cross sections for the individual $\chi_i \chi_j$ annihilations,
\beq
\langle \sigma_{ij} v\rangle =\sigma_{ij} x^{-n}
\eeq
\beq
\langle \sigma_{eff} v\rangle =\frac{\sum_{i,j=1}^N w_iw_j\sigma_{ij}x^{-n}}{
\left( \sum_{i=1}^N w_i \right)^2},~~~ w_i=
\left( \frac{m_i}{m_1}\right)^{3/2} e^{-x\left( \frac{m_i}{m_1}-1\right)} .
\label{omcopif}
\eeq
The relic abundance is now given by
\beq
\Omega h^2 =
\frac{8.7\times 10^{-11} \gev^{-2} }{\sqrt{g_*}
\int_{x_f}^\infty dx ~\langle \sigma_{eff} v\rangle x^{-2}}.
\label{omco}
\eeq
If all $N$ states are mass-degenerate, \eq{omco} takes the same form
as \eq{omsi} with the replacement $\sigma \to \sum_{i,j} \sigma_{ij}/N^2$

{\bf Bino}

The annihilation cross section of a pure $\bino$ into massless fermions is
\beq
\sigma_{\bino \bino}=\sum_f
\frac{g^4 t_W^4 (T_f-Q_f)^4r(1+r^2)}{2\pi m_{\tilde f}^2 x(1+r)^4}
, ~~~x\equiv \frac{M_1}{T}
,~~~r\equiv \frac{M_1^2}{m_{\tilde f}^2},
\eeq
where $T_f$ and $Q_f$ are third component of isospin and electric charge
of the fermion $f$. The value of $\Omega$ is obtained from \eq{omsi}.

{\bf Higgsino}

For the Higgsino we consider 4 coannihilating states with equal mass $\mu$:
$( \hino_1, \hino_2, \hino^+, \hino^- )$. The neutral states $\hino_{1,2}$
are defined as in sect.~\ref{secbh}, and we treat $\hino^+$ and $\hino^-$ as
independent to have the same number of degrees of freedom for each state.
The thermal-averaged cross sections for the individual annihilation
processes (assuming $\mu\gg M_W$ and heavy supersymmetric scalars)
is described by the symmetric matrix $\sigma_{ij}$ with
\beq
\sigma_{11}=\sigma_{22}=\frac{g^4}{128\pi\mu^2}\left( \frac 32 +t_W^2
+\frac{t_W^4}{2}\right)
\eeq
\beq
\sigma_{12}= \frac{g^4}{128\pi\mu^2}\left( \frac {29}4 +\frac{21}{2} t_W^4
\right)
\eeq
\beq
\sigma_{1+}=\sigma_{1-}=\sigma_{2+}=\sigma_{2-}=
 \frac{g^4}{128\pi\mu^2}\left( 6 +t_W^2
\right)
\eeq
\beq
\sigma_{+-}=\frac{g^4}{128\pi\mu^2}\left( \frac {29}4 +t_W^2
+11t_W^4\right)
\eeq
\beq
\sigma_{++}=\sigma_{--}=\frac{g^4}{64\pi\mu^2}
\eeq
The effective cross section $\langle \sigma_{eff}v\rangle
=\sum_{1,j=1}^4\sigma_{ij}/16$
and the Higgsino relic abundance are given in eqs.~(\ref{shig})--(\ref{ohig}).

{\bf Wino}

For the Wino we consider 3 coannihilating states with equal masses $M_2$:
$( \wino^0 , \wino^+,\wino^- )$. The symmetric matrix  $\sigma_{ij}$
(assuming $M_2\gg M_W$ and heavy supersymmetric scalars) is
given by
\beq
\sigma_{00}= \frac{g^4}{8\pi M_2^2}, ~~~ \sigma_{0+}=\sigma_{0-}=
\frac{7g^4}{32\pi M_2^2},\label{win1}
\eeq
\beq
\sigma_{+-}= \frac{9g^4}{32\pi M_2^2}, ~~~  \sigma_{++}=\sigma_{--}=
\frac{g^4}{16\pi M_2^2}. \label{win2}
\eeq
The effective cross section $\langle \sigma_{eff}v\rangle
=\sum_{1,j=1}^3\sigma_{ij}/9$
and the Wino relic abundance are given in eqs.~(\ref{swin})--(\ref{owin}).

{\bf Well-Tempered Neutralino}

Let us consider the case of the tempered $\bino$/$\wino$, where there are
4 coannihilating states: $( \bino , \wino^0 , \wino^+,\wino^- )$.
We compute the leading contributions to the relevant cross sections,
in the limit of unbroken electroweak symmetry and heavy supersymmetric
scalars. The entries of the matrix $\sigma_{ij}$ relative to Wino
states are given by eqs.~(\ref{win1})--(\ref{win2}), while the other
entries are given by
\beq
\sigma_{\bino \bino }= \sigma_{\bino \bino }^{(s)}+
\sigma_{\bino \bino }^{(p)}
\eeq
\beq
\sigma_{\bino \bino }^{(s)}=  \frac{g^4\theta^4 M_1^2}{2\pi (M_1^2+M_2^2)^2}
\eeq
\bea
\sigma_{\bino \bino }^{(p)}&=&
\frac{g^4t_W^4}{64\pi x(M_1^2+\mu^2)^4}
\left[ M_1^2 \left( 3 M_1^4 +2 M_1^2\mu^2 +3\mu^4 \right) \right.
\nonumber \\
&+& \left. 2M_1\mu \left( M_1^4 +4 M_1^2\mu^2 +3\mu^4 \right) s_{2\beta} +
3\mu^2 \left( M_1^2+\mu^2 \right)^2 s_{2\beta}^2 \right]
\label{spluga}
\eea
\beq
\sigma_{\bino \wino^0 }= \frac{g^4\theta^2}{8\pi M_1^2}
\left( 1 -4\Delta +\frac{31}{8} \Delta^2 \right) +{\cal O}\left( \Delta^3
\right)
\eeq
\beq
\sigma_{\bino \wino^+ }=\sigma_{\bino \wino^- }=
\frac{g^4\theta^2}{128\pi M_1^2}
\left( 28 -\frac{107}{2}\Delta +\frac{141}{2} \Delta^2 \right)
+{\cal O}\left( \Delta^3
\right) .
\eeq
The $\bino$/$\wino$ mixing angle $\theta$ and the mass-splitting parameter
$\Delta$ are defined in \eq{tett}.
$\sigma_{\bino \bino }^{(p)}$ does not vanish in the limit $\theta
\to 0$,
corresponding to the contribution from the Bino annihilation into Higgs bosons
and longitudinal gauge bosons via $t$-channel Higgsino exchange. This
process can only occur in $p$-wave, as exhibited by the factor $x=M_1/T$
in \eq{spluga}, but it becomes important in the limit of small
mixing angle. In \eq{spluga} we have neglected the lightest
Higgs mass and taken the pseudoscalar Higgs mass to be larger than $M_{1,2}$.
The value of the relic abundance, determined by
eqs.~(\ref{omcopif})--(\ref{omco}) is in excellent agreement with the
results obtained with DARKSUSY~\cite{darksusy}, in the region of small $\theta$ and
quasi-degenerate gauginos.

It is interesting to consider the limit of vanishing mixing angle $\theta$
(but not exactly zero, in order to allow equilibrium between $\bino$ and
$\wino$ number densities). The relic density of well-tempered $\bino$/$\wino$
neutralinos (with $|M_1|<|M_2|$) is completely determined by the $\wino$
annihilation processes (neglecting the small $p$-wave annihilation
into Higgs fields) and we find
\beq
 \Omega_{\bino \wino} h^2 =0.13 \left( \frac{M_2}{2.5 \tev}\right)^2
\frac{1}{R_{\wino}}
\label{omapp}
\eeq
\beq
R_{\wino}=
\int_0^1 dy \left[ 1+\frac{1}{3} \left( \frac{M_1}{M_2}
\right)^{3/2} e^{\frac{x_f}{y}\left( \frac{M_2}{M_1} -1\right) } \right]^{-2}
\simeq \left( \frac 34 \right)^2 e^{-\xi_{\wino} x_f
\left( \frac{M_2}{M_1} -1\right) },
\eeq
with $\xi_{\wino}\simeq 1.7$.
This result well describes the dependence of $ \Omega_{\bino \wino}$ on
the $\bino$/$\wino$ mass ratio.
Notice that the mixing angle $\theta$ is typically
rather small in most of the parameter space under consideration, and that
the dependence of the relic abundance on $\theta$ is modest (for small
$\theta$), see fig.~\ref{figbw}. Therefore \eq{omapp} is a good approximation
in most of the well-tempered $\bino$/$\wino$ region.

The expressions of the annihilating cross section in the case of well-tempered
$\bino$/$\hino$ are rather long and we do not show them here. However, the
result becomes particularly simple in the case of very small mixing
angles (neglecting the small $p$-wave annihilation
into Higgs fields),
where the relic abundance is determined by Higgsino annihilation.
In this case, we find
\beq
\Omega_{\hino} h^2 =0.10 \left( \frac{\mu}{1 \tev}\right)^2
\frac{1}{R_{\hino}}
\label{spit}
\eeq
\beq
R_{\hino}=
\int_0^1 dy \left[ 1+\frac{1}{4} \left( \frac{M_1}{\mu}
\right)^{3/2} e^{\frac{x_f}{y}\left( \frac{\mu}{M_1} -1\right) } \right]^{-2}
\simeq \left( \frac 45 \right)^2 e^{-\xi_{\hino} x_f
\left( \frac{\mu}{M_1} -1\right) },
\eeq
with $\xi_{\hino}\simeq 1.5$.
The effect of the mixing angle is more important for well-tempered
$\bino$/$\hino$ rather than $\bino$/$\wino$,
since $\theta_\pm$ is parametrically
larger than $\theta$ by a factor $\mu /M_Z$, see eqs.~(\ref{mix}) and
(\ref{tett}). For this reason, \eq{spit} has a more limited applicability.
However, in some cases, the $\bino$/$\hino$ mixing effects becomes negligible;
this happens, for instance, when $\mu$ and $M_1$ have opposite signs and
$\tan\beta$ is small.

In the limit in which the mass difference among neutralino states is
sufficiently large and coannihilation can be neglected, the
annihilation cross section for the well-tempered $\bino$/$\hino$ is
described by
\beq
\sigma_{\bino \bino }= \sigma_{\bino \bino }^{(s)}+
\sigma_{\bino \bino }^{(p)}
\eeq
\bea
\sigma_{\bino \bino }^{(s)}&=&
\frac{g^4\ \left(\theta_-^4+\theta_+^4\right)  M_1^2}{64\pi\left(
\mu^2+M^2_1\right)}\left( 3+
2t^2_W+t^4_W\right) \nonumber\\
&+&\frac{g^4\theta_-^2\theta_+^2}{128\pi}\left[
\frac{77 \mu^4-26\mu^2M^2_1-39M^4_1}{(\mu^2+M_1^2)^3}
+\frac{2}{M_1^2}\left( 7 t_W^4-6t_W^2+3\right) \right]
\eea
where $\sigma_{\bino \bino }^{(p)}$ is given in \eq{spluga}
and $\theta_\pm$ are defined in \eq{mix}.

\section*{Appendix B}

Here we give the analytic expressions of the well-tempered neutralino
masses and mixing angles in the case of complex soft terms.
In general the parameter $M_{1,2}$ and $\mu$ are expected to be
complex, although their
phases are constrained by measurements of electric dipole moments~\cite{edm}.
Maximal phases are allowed when squarks and sleptons are very
heavy~\cite{split3,edms}, as
in the case of Split Supersymmetry.

Consider the Higgs/Higgsino/gaugino sector. The theory has 4 (potentially)
complex parameters: $M_1$, $M_2$, $\mu$, and $B_\mu$. By redifining the
phases of $\bino$, $\wino$, $\hino_{u,d}$, and the two Higgs fields,
under the constraint of keeping real the
gauge coupling constants in the $H$--$\hino$--$\bino$ and
$H$--$\hino$--$\wino$ vertices, the complex parameters transform as
$B_\mu \to e^{i\gamma}B_\mu$, $\mu \to e^{i{\tilde \gamma}}\mu$,
$M_{1,2} \to e^{i({\gamma -\tilde \gamma})}M_{1,2}$. Therefore the invariants
under phase reparametrization are ${\rm arg} (M_1^*M_2)$ and
${\rm arg}(B_\mu^* \mu M_{1,2)}$, but only 2 out of 3 are independent.
Since after the minimization of the Higgs potential the phase of $B_\mu$
corresponds to the phase of $s_{2\beta}$, we choose an independent phases
\beq
\varphi_1\equiv {\rm arg}(s_{2\beta}\mu M_1 ) ,~~~
\varphi_2\equiv {\rm arg}(s_{2\beta}\mu M_2 ) .
\label{fasi}
\eeq
In the following, we choose to work in the standard basis where
both Higgs vacuum expectation values are real. Notice that, in the limit
of large $\tan\beta$, the invariants in \eq{fasi} (and the
dependence on the sign of $\mu$) vanish, and only the
relative gaugino-mass phase ${\rm arg} (M_1^*M_2)$ is physical.

{\bf Well-Tempered Bino/Wino}

At the leading order in the $1/\mu$ expansion, the unitary matrix $N$
that diagonalizes the neutralino mass matrix in the $\bino$/$\wino$ sector,
according to $N {\cal M} N^T$, is given by
\beq
N=D\pmatrix{1 & -\theta \cr \theta^* & 1},~~
\theta = \frac{s_{2W}s_{2\beta} M_Z^2\left(M_1 \mu +M_2^*\mu^* \right)}
{2|\mu|^2 \left( |M_2|^2-|M_1|^2\right)}  .
\label{cip}
\eeq
The diagonal matrix $D_{ij}=\delta_{ij} \sqrt{|m_{\chi_i}|/m_{\chi_i}}$
removes the phases of the mass eigenvalues, to make them real and positive.
The absolute values of the mass eigenvalues for neutral and charged states are
\beq
|m_{\chi_1}|=|M_1|-\frac{s_W^2s_{2\beta}c_{\varphi_1}M_Z^2}{|\mu|} ,~~~
|m_{\chi_2}|=|m_{\chi^+}|=
|M_2|-\frac{c_W^2s_{2\beta}c_{\varphi_2}M_Z^2}{|\mu|} .
\label{ciop}
\eeq
Notice how the
dependence on gaugino-mass phases
always appears in the form of the invariants under phase reparametrization
defined in \eq{fasi}.
Equations~(\ref{cip})--(\ref{ciop})
interpolate between the two cases of equal and opposite
signs for the gaugino masses, considered in the text.
In particular, near the well-tempered condition, \eq{cip} gives
\beq
|\theta |= \frac{s_{2W}s_{2\beta} M_Z^2}
{2|\mu| \left( |M_2|-|M_1|\right)} \sqrt{ \frac{1+c_{\varphi_1+\varphi_2}}{2}}.
\eeq
This shows that $|\theta|$ vanishes only in the particular case in which
$M_1$ and $M_2$ are real and have opposite sign.

{\bf Well-Tempered Bino/Higgsino}

At the leading order, the unitary matrix $N$
that diagonalizes the neutralino mass matrix according to $N {\cal M}
N^T$, in the $\bino$/$\hino$ sector is
\beq
N=D\pmatrix{1 & \theta_+ & \theta_- \cr
-e^{-i\phi_+} \theta_+^* & \cos(\alpha) & i \sin(\alpha) \cr
-e^{-i\phi_-} \theta_-^* & i\sin(\alpha) & \cos(\alpha)}
\eeq
\beq
\theta_\pm =\frac{s_W M_Z \left( s_\beta \pm c_\beta \right)
\left( \mu^* \pm M_1 \right)}
{\sqrt{2} \left( |\mu|^2-|M_1|^2\right) }
\label{sppp1}
\eeq
\beq
\tan \alpha =\frac{2{\rm Im}\left( \theta_+^*\theta_-
\right)}{|\theta_+|^2-|\theta_-|^2},~~~~
\sin \phi_\pm = -\sin\alpha~{\rm Re} \frac{\theta_\mp}{\theta_\pm} .
\label{sppp2}
\eeq
The diagonal matrix $D$ is defined as above, and the absolute values of the
mass eigenvalues are
\beq
|M_1| -s_W^2M_Z^2 \left( \frac{|M_1|+|\mu|s_{2\beta}c_{\varphi_1}}{|\mu|^2
-|M_1|^2}\right) ,
\label{auto1}
\eeq
\beq
|\mu| +\frac{s_W^2M_Z^2}{2\left( |\mu|^2-|M_1|^2 \right)}
\left[ |\mu|+|M_1|s_{2\beta}c_{\varphi_1}
\pm \sqrt{ \left( |\mu| s_{2\beta}+|M_1|c_{\varphi_1}\right)^2
+ |M_1|^2c_{2\beta}^2 s_{\varphi_1}^2}
\right]
\label{auto2}
\eeq

Notice an important difference with the case of real soft parameters.
The states $\hino_{1,2}$ defined in sect.~\ref{secbh} are not
approximate mass eigenstates, for a generic phase $\varphi_1$,
but they are rotated by an angle $\alpha$,
which is typically of order unity. In the particular cases $\varphi_1=0,\pi$
(corresponding to real $M_1$ and $\mu$, with equal or opposite signs),
$\theta_\pm$ are real, and $\alpha =\phi_\pm =0$, see
eqs.~(\ref{sppp1})--(\ref{sppp2}). In this case the mixing between
 $\hino_1$ and $\hino_2$ appears only at ${\cal O}(M_Z^2)$, as shown
in \eq{nrot}.
The mass eigenvalues
given in eqs.~(\ref{auto1})--(\ref{auto2}) reduce to
\beq
|M_1| -s_W^2M_Z^2 \left( \frac{|M_1|+|\mu|s_{2\beta}c_{\varphi_1}}{|\mu|^2
-|M_1|^2}\right) ,~~~
|\mu|+\frac{ s_W^2M_Z^2 (1\pm s_{2\beta})}{2\left( |\mu|
\mp c_{\varphi_1} |M_1|\right)},
\eeq
where $c_{\varphi_1}=\pm 1$, when $M_1$ and $\mu$ have
equal or opposite signs, respectively. This result
corresponds to \eq{pix}.


\end{document}